\documentclass[aps,reprint,superscriptaddress,shortbibliography,prl]{revtex4-1}

\usepackage{graphicx, amsmath, verbatim, dsfont, amsfonts, color}
\usepackage{braket}
\usepackage[us]{datetime}

\begin{document}

\title{Mesoscopic transport in electrostatically-defined spin-full channels in quantum Hall ferromagnets}

\author{Aleksandr~Kazakov}
\affiliation{Department of Physics and Astronomy, Purdue University, West Lafayette, IN 47907 USA}
\author{George~Simion}
\affiliation{Department of Physics and Astronomy, Purdue University, West Lafayette, IN 47907 USA}
\author{Yuli~Lyanda-Geller}
\affiliation{Department of Physics and Astronomy, Purdue University, West Lafayette, IN 47907 USA}
\affiliation{Birck Nanotechnology Center, Purdue University, West Lafayette, IN 47907 USA}
\author{Valery~Kolkovsky}
\affiliation{Institute of Physics, Polish Academy of Sciences, Al. Lotnikow 32/46, 02-668 Warsaw, Poland}
\author{Zbigniew~Adamus}
\affiliation{Institute of Physics, Polish Academy of Sciences, Al. Lotnikow 32/46, 02-668 Warsaw, Poland}
\author{Grzegorz~Karczewski}
\affiliation{Institute of Physics, Polish Academy of Sciences, Al. Lotnikow 32/46, 02-668 Warsaw, Poland}
\author{Tomasz~Wojtowicz}
\affiliation{Institute of Physics, Polish Academy of Sciences, Al. Lotnikow 32/46, 02-668 Warsaw, Poland}
\affiliation{International Research Centre MagTop, al. Lotnikow 32/46, PL 02-668 Warszawa, Poland}
\author{Leonid~P.~Rokhinson}
\email{leonid@purdue.edu}
\affiliation{Department of Physics and Astronomy, Purdue University, West Lafayette, IN 47907 USA}
\affiliation{Birck Nanotechnology Center, Purdue University, West Lafayette, IN 47907 USA}
\affiliation{Department of Electrical and Computer Engineering, Purdue University, West Lafayette, IN 47907 USA}

\begin{abstract}
In this work we use electrostatic control of quantum Hall ferromagnetic transitions in CdMnTe quantum wells to study electron transport through individual domain walls (DWs) induced at a specific location. These DWs are formed due to hybridization of two counter-propagating edge states with opposite spin polarization. Conduction through DWs is found to be symmetric under magnetic field direction reversal, consistent with the helical nature of these DWs. We observe that long domain walls are in the insulating regime with localization length 4 - 6~$\mu$m. In shorter DWs the resistance saturates to a non-zero value at low temperatures. Mesoscopic resistance fluctuations in a magnetic field are investigated. The theoretical model of transport through impurity states within the gap induced by spin-orbit interactions agrees well with the experimental data. Helical DWs have required symmetry for the formation of synthetic p-wave superconductors. Achieved electrostatic control of a single helical domain wall is a milestone on the path to their reconfigurable network and ultimately to a demonstration of braiding of non-Abelian excitations.
\end{abstract}

\maketitle

The prediction that one-dimensional (1D) wires with lifted Kramers degeneracy but preserved time reversal symmetry coupled to a conventional superconductor can harbor non-Abelian excitations \cite{Kitaev2001} motivated development of various systems with conducting 1D helical channels. The required symmetry has been predicted \cite{Lutchyn2010,Oreg2010} and demonstrated in nanowires with strong spin-orbit interaction in the presence of a magnetic field \cite{Rokhinson2012,Das2012,Churchill2013}, at the edges of the quantum spin Hall effect devices \cite{Konig2007}, and in atomic chains with helical magnetic structure \cite{Nadj-Perge2014}. None of the aforementioned systems are easily reconfigurable, which hinders demonstration of braiding of quasiparticles and non-Abelian statistics. 

Edge states in the quantum Hall effect (QHE) regime have been used as a canonical system to study 1D Luttinger liquids \cite{Wen1990}, which are \textit{chiral} and not time reversal invariant. However, there is one overlooked regime in the QHE, quantum Hall ferromagnetic (QHFm) transition -- where \textit{helical} channels can be formed. Spin polarization of the topmost Landau level is determined by a competition between Zeeman, cyclotron, and exchange energies. Changing the balance between these energies (e.g. by applying an in-plane magnetic field) can lead to a QHFm transition where a uniform 2D gas spontaneously phase-separates into regions with different spin polarizations \cite{Verdene2007}. Domain walls at the boundaries of insulating ferromagnetic domains form \textit{helical} 1D channels (hDWs) \cite{Falko1999,Jungwirth2001,Chalker2002,Brey2002,Mitra2003}, and transport through a random network of conducting DWs has been studied in a context of a 2D phase transition \cite{DePoortere2000,Jaroszynski2002,Teran2010}. In the past, the study of an individual hDW was not feasible. In this paper we use recently developed gate control of the QHFm transition \cite{Kazakov2016} in CdMnTe quantum wells to demonstrate that hDWs can be formed at a specific location using electrostatic gating and we also present investigation of transport properties of isolated hDWs.

\begin{figure}
\centering\includegraphics[width=0.9\columnwidth]{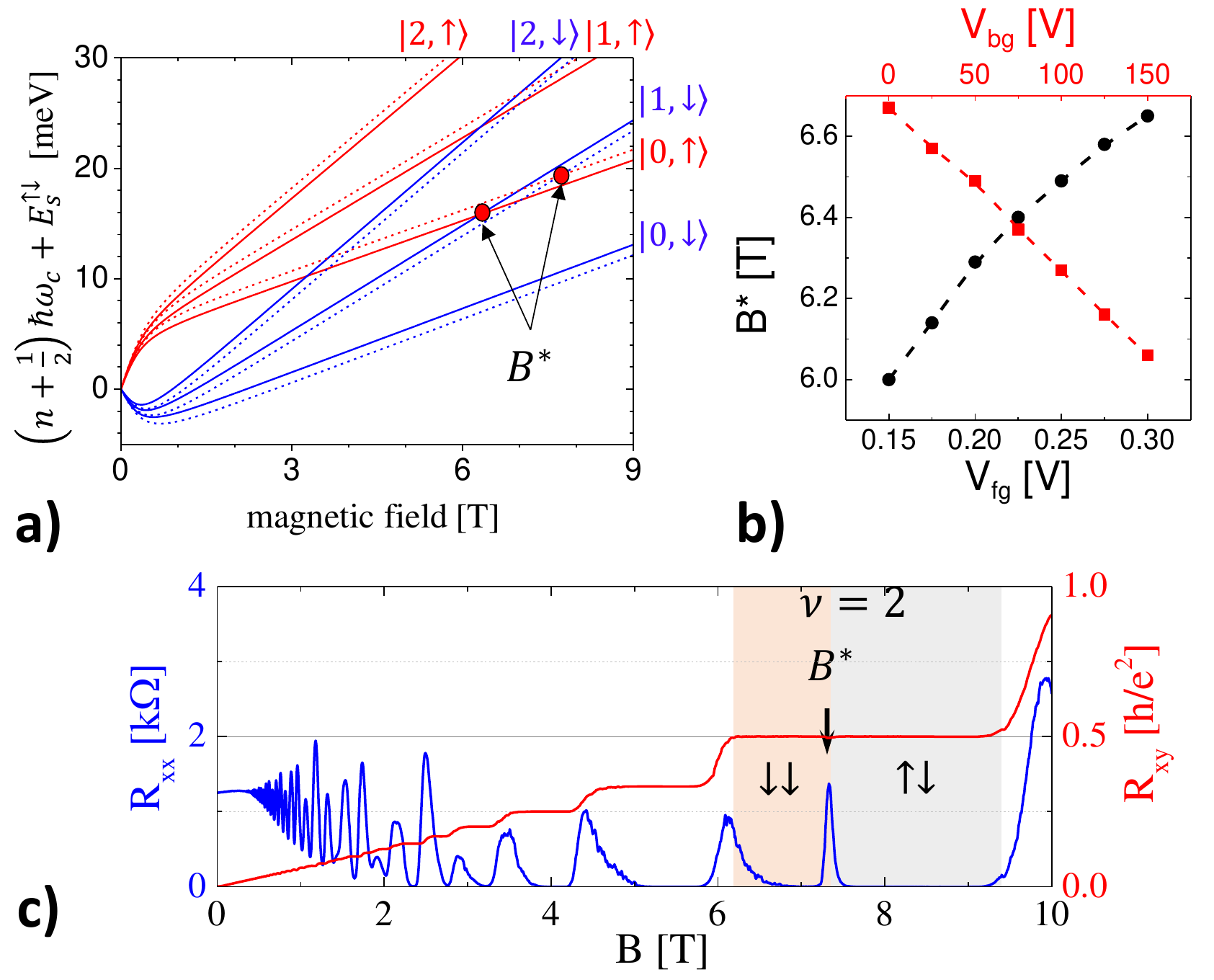}
\vspace{0in}
\caption{(a) Calculated energy spectrum of Landau levels in CdMnTe with 1.7\% Mn doping and s-d overlap $0.9\chi_0$ (solid lines) and $1.1\chi_0$ (dashed lines), where $\chi_0$ is the overlap for zero gate voltages. $B^*$ marks QHFm transitions where the ground state changes from $\ket{0\uparrow}$ to $\ket{1\downarrow}$. (b) Experimentally measured shift of QHFm at $\nu=2$ as a function of front ($V_{fg}$) and back ($V_{bg}$) gate voltages. (c) Longitudinal and Hall resistance measured at an elevated temperature 300~mK. Sharp peak at 7.3~T within the $\nu=2$ shaded region is a QHFm transition between fully polarized and unpolarized states, where the top filled Landau level changes polarization.}
\label{effZ}
\vspace{-0.2in} 
\end{figure}

The QHFm transition was first observed at a filling factor $\nu=2/3$ \cite{Eisenstein1990} in high mobility GaAs quantum wells. In this paper we focus on the QHFm transition at $\nu=2$ in CdMnTe dilute magnetic semiconductor quantum wells \cite{Wojtowicz2000}, where QHFm transitions in both integer \cite{Jaroszynski2002} and fractional \cite{Betthausen2014} QHE regimes have been observed. The QHFm transition in CdMnTe originates from a competition between negative Zeeman energy (the Land\'e g-factor of CdTe $g=-1.6$) and positive exchange energy between s-electrons in the quantum well and d-shell electrons in Mn. Presence of the s-d exchange modifies Landau levels, see Fig.~\ref{effZ}a, and can result in the crossing of levels with different polarizations at high magnetic fields. The magnetic field $B^*$ corresponds to a cancellation of differences in total Zeeman, cyclotron, and exchange energies of $\ket{0\uparrow}$ and $\ket{1\downarrow}$ states. At this field levels would cross, but spin-orbit interaction introduces a small avoided crossing \cite{Kazakov2016}. When driven through $B^*$, the 2D gas undergoes a polarized ($\downarrow\downarrow$) to unpolarized ($\downarrow\uparrow$) phase transition at $\nu=2$ (only two Landau levels filled), which is marked on the plot. We observe that in transport this QHFm transition is seen as a sharp peak in the longitudinal resistance in the middle of the $\nu=2$ plateau, Fig.~\ref{effZ}c.

Electrostatic control of the QHFm transition in CdMnTe was developed in \cite{Kazakov2016}, where we introduced non-uniform placement of Mn in the growth direction within the quantum well. The electric field shifts the electron wavefunction relative to the Mn position thereby controlling the s-d overlap $\chi(V_g)$. The corresponding change in the strength of the s-d exchange results in the shift of $B^*$, as shown for two values of $\chi$ in Fig.\ref{effZ}a. Experimentally we can control $B^*$ within $\sim 10\%$ by both front and back gates as shown in Fig.~\ref{effZ}b.
 
\begin{figure}
\centering\includegraphics[width=0.9\columnwidth]{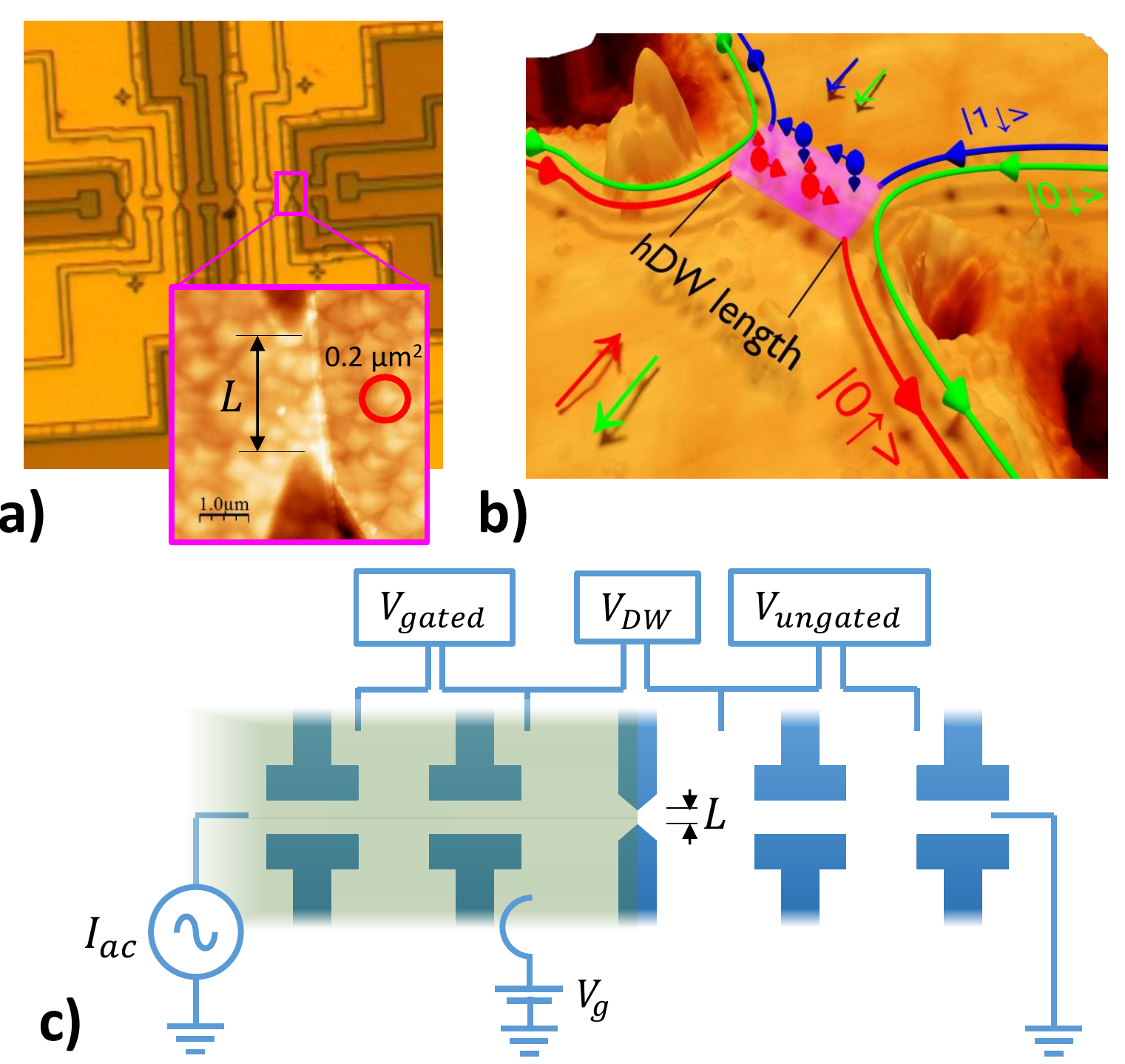}
\vspace{0in}
\caption{(a) Optical image of a sample, dark areas are etched and yellow areas are covered by a top gate. Inset is an AFM image of a constriction, where the vertical gate boundary is clearly seen. (b) An artistic rendering of an AFM image at $\nu=2$ with a schematic flow of $\ket{0\uparrow}$, $\ket{0\downarrow}$, and $\ket{1\downarrow}$ edge channels assuming that the QHFm transition is gate-tuned across the constriction. $\ket{0\uparrow}$ and $\ket{1\downarrow}$ states hybridize forming a helical domain wall. (c) Schematic of the measurement setup.}
\label{f:samp}
\vspace{-0.2in} 
\end{figure}

Devices were fabricated from CdMnTe/Cd$_{0.8}$Mg$_{0.2}$Te QW heterostructures grown by molecular beam epitaxy, see Refs. \cite{Jaroszynski2002,Betthausen2014} for details. The QW is 30~nm wide and is modulation doped with iodine. Mn is introduced into the QW as 7 $\delta$-doping layers spaced by 6 monolayers of CdTe starting 13 nm from the bottom of the quantum well in the growth direction. The effective Mn concentration is $1.5\%-1.7\%$ as determined from the position of the QHFm transition $B^*$ at $\nu=2$. Low temperature density and mobility in ungated samples are $3-3.5\cdot10^{11}$~cm$^{-2}$ and $3-4\cdot10^4$~cm$^2$/V$\cdot$s respectively. The transition field $B^*$ at zero gate voltage can be adjusted by varying conditions of the LED illumination during a cooldown \cite{sup}. We attribute this tunability to different dopant ionization profiles and, consequently, different profile of the electron wavefunction within the quantum well. A semi-transparent front gate is formed by evaporating 10-15~nm of Ti on the surface of the sample, and a copper foil glued to the back of the sample serves as a back gate. Ohmic contacts are produced by soldering freshly cut indium pellets similar to previous studies \cite{Jaroszynski2002, Betthausen2014}. Electron transport is measured in a dilution refrigerator in a temperature range $30-650$~mK with a standard ac technique using excitation current $I_{ac}\leq1~nA$.

Samples are patterned in a number of gated and ungated Hall bar sections with sizes of $25~\mu$m length and $15~\mu$m width, see Fig.~\ref{f:samp}. The front gate boundary is aligned with narrow constrictions of various lithographical widths $L=1-15$~$\mu$m. The constrictions electrical width is reduced by $2l_D=200-400$~nm, where $l_D$ is depth of electrical depletion of a 2D electron gas near the mesa edges. It is further reduced by $\approx1.8l_D-2.5\sqrt{a_B\cdot l_D}=120-280$~nm ($a_B=5.4$ nm is the Bohr radius in CdTe) due to the formation of edge channels in the QHE regime \cite{Chklovskii1992}. The overall reduction is $0.5-1$~$\mu$m compared to the lithographic $L$. This sample design allows simultaneous measurement of longitudinal resistance $R_{xx}=V_{xx}/I_{ac}$ in gated ($R_{gated}$) and ungated ($R_{ungated}$) regions, as well as longitudinal resistance in the presence of the domain wall $R_{DW}$, Fig.\ref{f:samp}c. 

\begin{figure}
\centering\includegraphics[width=0.9\columnwidth]{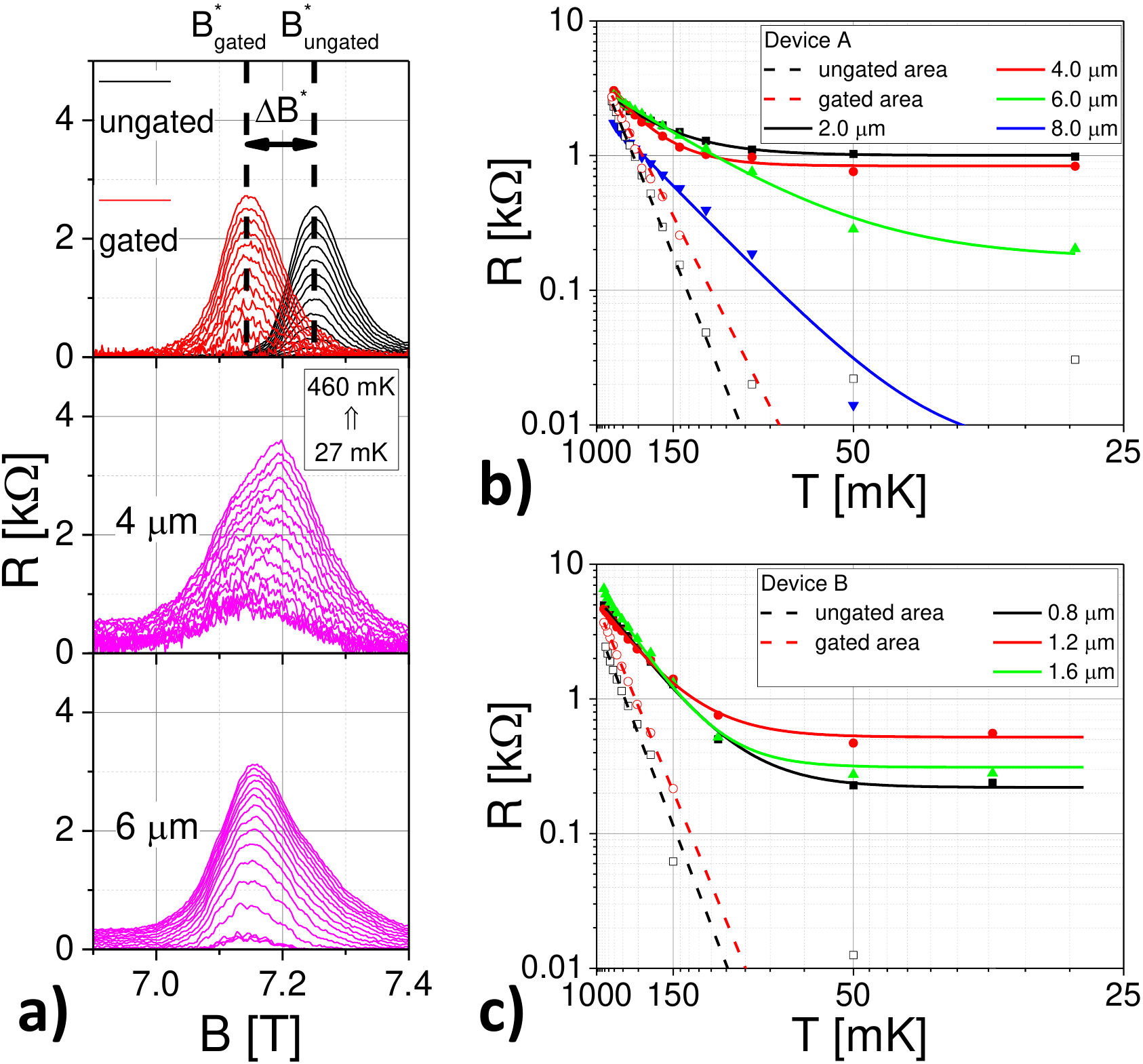}
\vspace{0in}
\caption{(a) The upper pane shows QHFm transitions for large ungated and gated areas. $R_{DW}$ for $L=4$ and 6~$\mu$m constrictions is plotted in the lower panes. Dashed lines mark $B^*$ in gated and ungated areas. At low temperatures the resistance of the $L=6\ \mu$m constriction almost vanishes, while for $L=4\ \mu$m saturates to a non-zero value. (b,c) $R_{DW}(T)$ dependence for constrictions with different $L$ for $\Delta B^*=0.11$~T in device A and $\Delta B^*=0.25$~T in device B are plotted in Arrhenius plot. Solid lines are fits to $R=R_0+A\cdot e^{-E_a/kT}$, dashed lines are fits to thermally activated conduction with a gap $\approx1$~K.}
\label{tdep}
\vspace{-0.2in} 
\end{figure}

The difference between QHFm transitions in gated and ungated regions is $\Delta B^*=B^*_{ungated}-B^*_{gated}$ and positions of $B^*$ within $\nu=2$ plateaus can be adjusted by a combination of cooldown conditions and gate voltages \cite{sup}. Note that the energy gap in the vicinity of the QHFm transition is $\sim\hbar e |\Delta B^*|/2m=\hbar\Delta\omega_c/2\approx0.57$~meV/T and increases with separation $\Delta B^*$. The value $\Delta B^*$ controls the gradient of the s-d exchange and the width of the hDWs, and can be adjusted between 0 and 0.3 T in our experiments. 

Magnetoresistance in the vicinity of the QHFm transition for $\Delta B^*=0.11$ T, where both $B^*_{gated}=7.14$~T and $B^*_{ungated}=7.25$~T are tuned into the middle of the $\nu=2$ plateau, is plotted in Fig.~\ref{tdep}. Here $\nu=2$ extends between $6.7$~T and $8.2$~T. $R_{xx}=0$ below 7.0~T corresponds to a fully polarized ($\downarrow\downarrow$) state with the $\ket{1\downarrow}$ topmost energy level filled, while $R_{xx}=0$ above 7.4~T is an unpolarized ($\downarrow\uparrow$) state with the topmost energy level $\ket{0\uparrow}$. Resistance of the QHFm transition peak for wide 2D regions shows activation behavior with an energy gap $\approx 1$~K, see dashed lines in Fig.~\ref{tdep}(b,c) attributed to spin-orbit coupling of $\ket{1\downarrow}$ and $\ket{0\uparrow}$ Landau levels \cite{Kazakov2016}. The value of $\Delta B^*$ is large enough that resistance in the midpoint $B=7.195$~T vanishes at low $T<100$~mK. Thus, The QHFm transition at a gate boundary should occur in the range $7.14$~T$<B<7.25$~T. Indeed, $R_{DW}$ peaks within that field range as shown in the middle and bottom panels in Fig.~\ref{tdep}a. For narrow (short) constrictions, $L<6\ \mu$m, resistance saturates at low temperatures to a non-zero value, see Fig.~\ref{tdep}(b,c). It is important to note that for $T<100$~mK the contribution of the wide 2D regions to $R_{DW}$ is negligible, and $R_{DW}$ originates from the conduction through the channel formed along the gate boundary.

\begin{figure}
\centering\includegraphics[width=0.95\columnwidth]{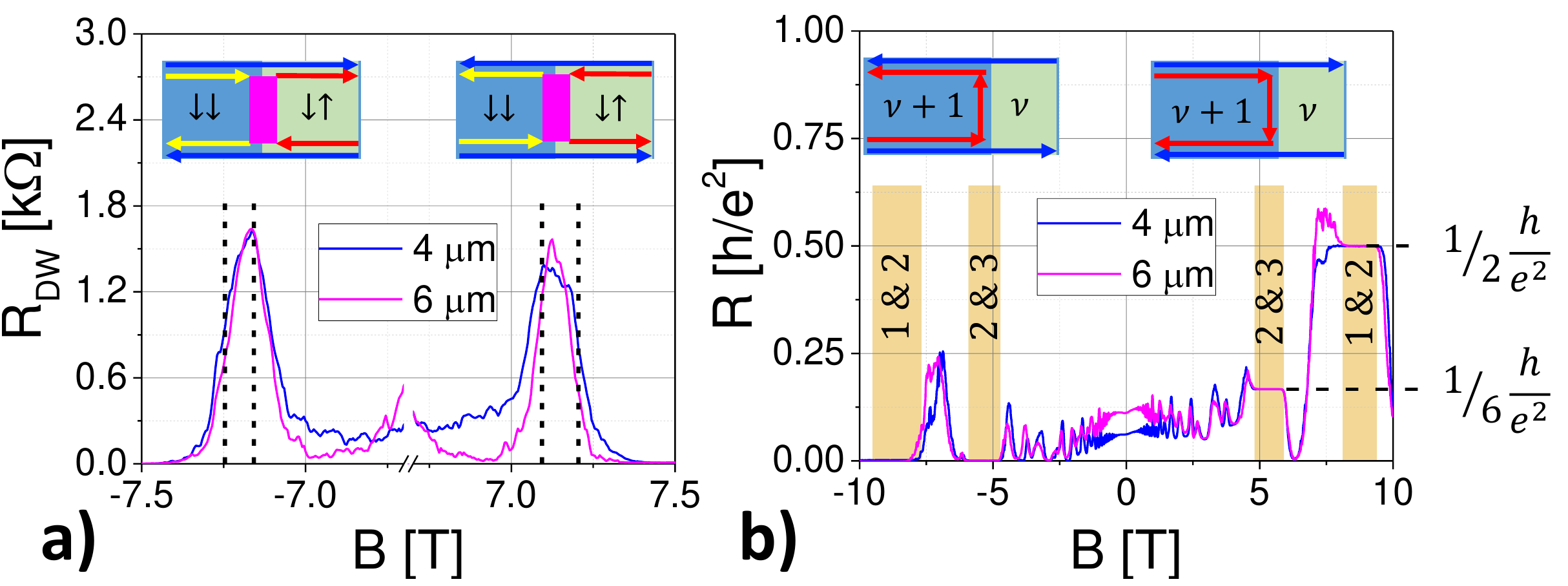}
\caption{(a) The resistance of the hDW is symmetric under magnetic field reversal. (b) In the presence of chiral channels formed at a boundary between two different QHE states, the resistance is highly asymmetric under magnetic field reversal (highlighted regions are for the boundaries between $\nu=1$ \& 2 and 2 \& 3 QHE states).}
\label{bdep}
\vspace{-0.2in} 
\end{figure}

One of the hallmarks of time reversal invariant \textit{helical} DWs is the symmetry with respect to magnetic field reversal, because domain walls emerge from two counterpropagating edges with the same filling factor. Indeed, we observed that $R_{DW}(B)\approx R_{DW}(-B)$, see Fig~\ref{bdep}a. This magnetic field reversal symmetry is in a striking contrast to properties of $R=R_{ch}$ measured when a \textit{chiral} channel is formed at a boundary of $\nu$ and $\nu+1$ QHE states, where $R=R_{ch}=0$ for one field direction and $R=R_{ch}=[\frac{1}{\nu}-\frac{1}{\nu+1}]^{-1}$ $h/e^2$ for the other direction \cite{Haug1993}. Indeed, at positive B, we see $R_{ch}=h/2e^2$ at the boundary between $\nu=1$ and $\nu=2$, and $R_{ch}=h/6e^2$ at the boundary between $\nu=2$ and $\nu=3$. However, with reversed B at the same boundaries $R_{ch}=0$, Fig~\ref{bdep}b.

\begin{figure}
\centering\includegraphics[width=\columnwidth]{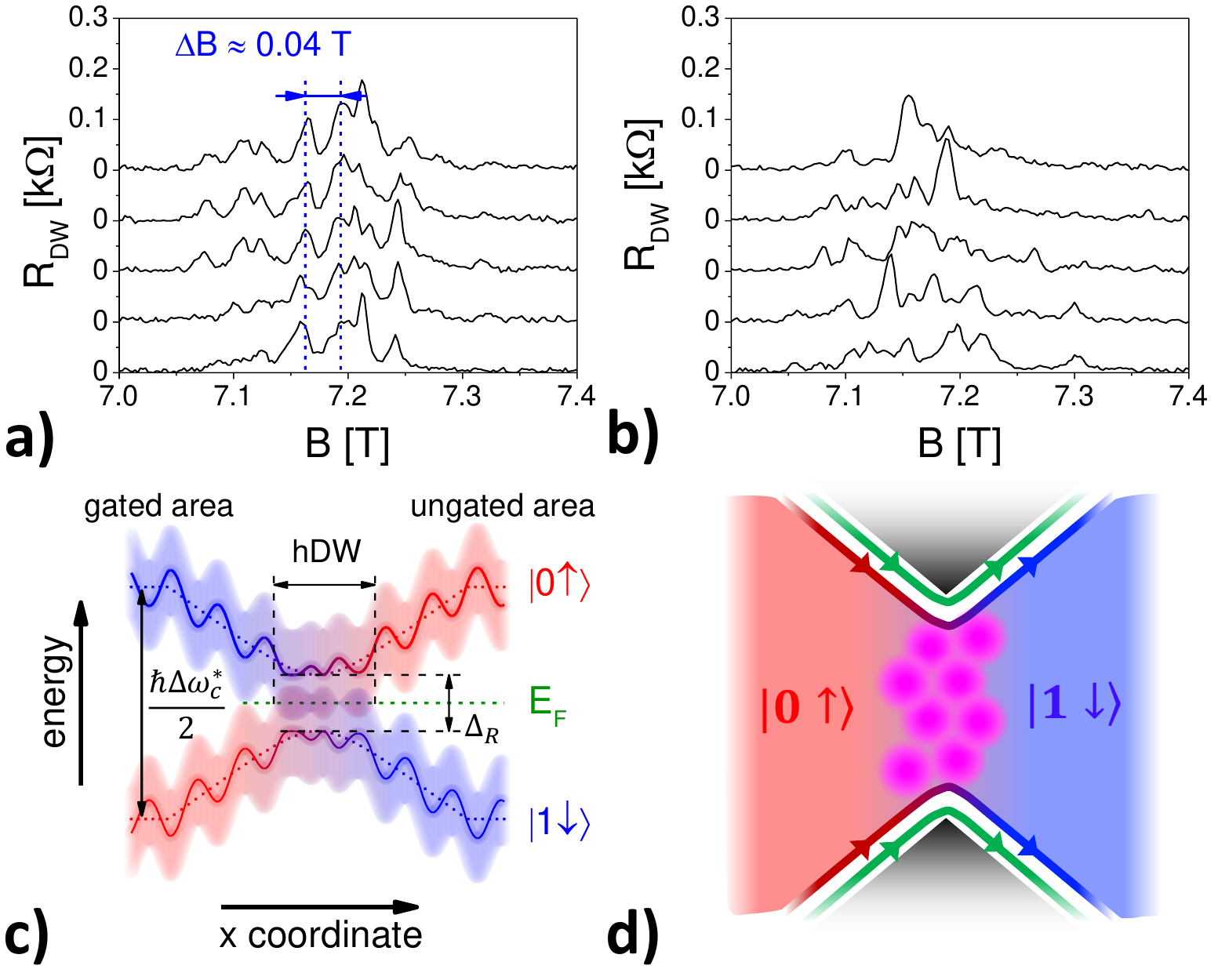}
\vspace{0in}
\caption{(a,b) Mesoscopic fluctuations measured in a device with $L=2\ \mu$m constriction at $T=27$~mK. In (a) magnetic field was swept within the $\nu=2$ state $6.8-7.5$~T. Fluctuations have a similar pattern with a quasi-period of $\Delta B\sim 40$~mT. In (b) $B$ was changed in a wide range $5-10$~T and the fluctuation pattern changes drastically. (c) Energy diagram of a hDW formed at the gate boundary. Wiggling lines indicate schematically a role of disorder and shaded areas are localized states in the tails of Landau levels. At low temperatures, conduction occurs via localized states in the gap. (d) Schematic of a conducting channel formed by coupled $\nu=2$ edge states. Electron tunneling via magenta in-gap states provide several interfering trajectories resulting in mesoscopic fluctuations of resistance.}
\label{ucf}
\end{figure}

Helical domain walls are formed by two counterpropagating edge channels along the gate boundary with opposite spin orientations. The measured values of $R_{DW}<1$~k$\Omega$. This demonstrates that counterpropagating edge channels at the same $\nu$ cannot be in the regime of ballistic transport as this would result in $R_{DW}= h/2e^2=12.9$~k$\Omega$, inconsistent with experimental observations. In order to quantify transport characteristics of hDW we describe them as resistors $r$ which connect $\nu=2$ edge states on the opposite sides of a constriction, as shown schematically in Fig.~\ref{f:samp}b. The resistance $r$ is defined by the voltage drop along the length of the domain wall as current flows in the same direction. This direction is perpendicular to the direction of change of spin polarization caused by the electrostatic gate (Fig. 2). Assuming there is no equilibration between $\nu=1$ and $\nu=2$ edge channels, within the Landauer-B\"uttiker formalism we obtain $R_{DW}=1/(4r+6)$ \cite{sup}. For all $r$, in this model $R_{DW}<1/6\ h/e^2=4.3$~k$\Omega$, consistent with measured values of $R_{DW}$.

Certain insight into the nature of the electronic transport through hDWs can be obtained from mesoscopic fluctuations observed at low temperatures. As shown in Fig.~\ref{ucf}a, in short hDWs quasi-periodic conductance fluctuations are clearly seen. The quasi-period $\Delta B$ of these oscillations is $\sim 40-55$~mT. Similar quasi-periodic resistance fluctuations were observed in mesoscopic devices for transitions between neighboring quantum Hall states \cite{Jain1988,Simmons1991}. From exponential decay of the fluctuation's amplitude we estimate the phase coherence length $l_{\phi}\propto T^{-1}\sim1-2\ \mu$m at base temperature \cite{sup}, comparable with the length of the hDWs. One possible interpretation of mesoscopic fluctuations is the formation of a multi-domain structure with a small network of hDWs spanning across the constriction. This seems unlikely. On the one hand, some static disorder, such as Mn doping fluctuations, potential fluctuations due to remote impurities, or surface roughness with characteristic size of 0.2 $\mu$m (see atomic force micrograph of the device surface in Fig.~\ref{f:samp}a) which results in fluctuation of the perpendicular component of the magnetic field, may act as pinning centers for domain formation. On the other hand, experimentally we found that the fluctuation pattern changes drastically every time the magnetic field is ramped outside the $\nu=2$ state, Fig.~\ref{ucf}b, which means that dynamic fluctuations rather than static impurities define the conduction path within the channel. This conclusion is further supported by the observation that the fluctuation pattern slowly changes over several hours even if the field is kept close to the QHFm transition \cite{sup}. We also note that the width of the gate-defined potential gradient, which coincides with the region of s-d exchange gradient and defines the width of the conductive channel, is of the order of the 2D gas-to-gate distance ($\approx100$ nm), similar to the expected width of hDWs defined by the spin-orbit coupling and a gradient of exchange interaction. Thus, formation of a multi-domain structure is highly unlikely. Assuming the width of the hDW to be $\sim100$~nm, the period of quasi-periodic oscillations is close to the area of a single hDW formed in a $L=2\ \mu$m constriction.

In long channels $L>6\ \mu$m we observed suppression of conduction at low temperatures. Similarly to the bulk Landau levels, edge states with spins $\ket{0\uparrow}$ and $\ket{1\downarrow}$ do not cross and exhibit a spin-orbit gap $\Delta_R\approx50\ \mu$eV. Electron states in the gap in long channels become localized, i.e., strongly decay on the scale of the length of constriction $L>6\ \mu$m. Thus transport thermally activated over the gap is a dominant mechanism of conduction in such channels. In short $L<4\ \mu$m hDWs, in-gap states should provide a conduction path at low temperatures. The in-gap states are due to charge defects binding electrons in the tail of Landau levels. This is consistent with the experimental observation that large changes in magnetic field (i.e., the shift of Landau levels relative to the Fermi energy) alter the interference pattern. This model is visualized in Fig.~\ref{ucf}(c), where anticrossing of broadened Landau levels with in-gap states at the Fermi level is shown schematically to form a single hDW. Within this picture transport through a single hDW can be modeled numerically, see Supplemental Material \cite{sup} for details. In the model we assume that the primary source of localized states in the spin-orbit gap are potential fluctuations due to the remote doping, and we use the zero-field mobility to calculate the strength and density of the fluctuations. We also include surface roughness, which leads to the deviation of magnetic field orientation and orientation of Mn spins from the $z-$direction at high fields, effectively introducing a magnetic disorder. The calculated conductance of a hDW is $1/r=0.146\pm0.026\ e^2/h$ which corresponds to $R_{DW}=0.66-0.87$~$k\Omega$, in good agreement with experiment. Modeling also confirms that transport is indeed dominated by the conduction via in-gap states. The calculated hDW resistance yields resistance $R_{DW}$ symmetric under magnetic field reversal.

In conclusion, we demonstrated a conducting helical domain wall electrostatically defined at a designed location and studied its transport properties. We have found that long $L>6$~$\mu$m hDWs are insulating at low temperatures, consistent with the activation behavior of the QHFm transition in the bulk. Short $L<6$~$\mu$m hDWs remain conducting even at low temperatures. We find that conduction in short hDWs occurs via in-gap states, and conduction is symmetric under magnetic field reversal, a hallmark of helical channels. These hDWs, coupled to a s-wave superconductor, should support non-Abelian excitations. The investigated electrostatic control of hDW formation and transport provide a mechanism to form a reconfigurable network of helical channels and is an important milestone toward realization of braiding of non-Abelian excitations.

\begin{acknowledgments}
Authors acknowledge support by the Department of Defense Office of Naval research Award N000141410339 (A.K, T.W., G.S., Y. L-G. and L.R.) and by the National Science Centre (Poland) grant DEC-2012/06/A/ST3/00247 (V.K., Z.A., G.K., and T.W.) and by the Foundation for Polish Science through the IRA Programme financed by EU within SG OP Programme (V.K., Z.A., G.K., and T.W.).
\end{acknowledgments}

%


\clearpage
\newpage

\renewcommand{\thefigure}{S\arabic{figure}}
\renewcommand{\theequation}{S\arabic{equation}}
\renewcommand{\thepage}{sup-\arabic{page}}
\setcounter{page}{1}
\setcounter{equation}{0}
\setcounter{figure}{0}

\begin{center}
\textbf{\Large Supplementary Materials} \\
\vspace{0.2in} \textsc{Mesoscopic transport in electrostatically-defined spin-full channels in quantum Hall ferromagnets}\\
\end{center}

\section{AFM images of samples}

\begin{figure}[h]
\centering\includegraphics[width=0.7\columnwidth]{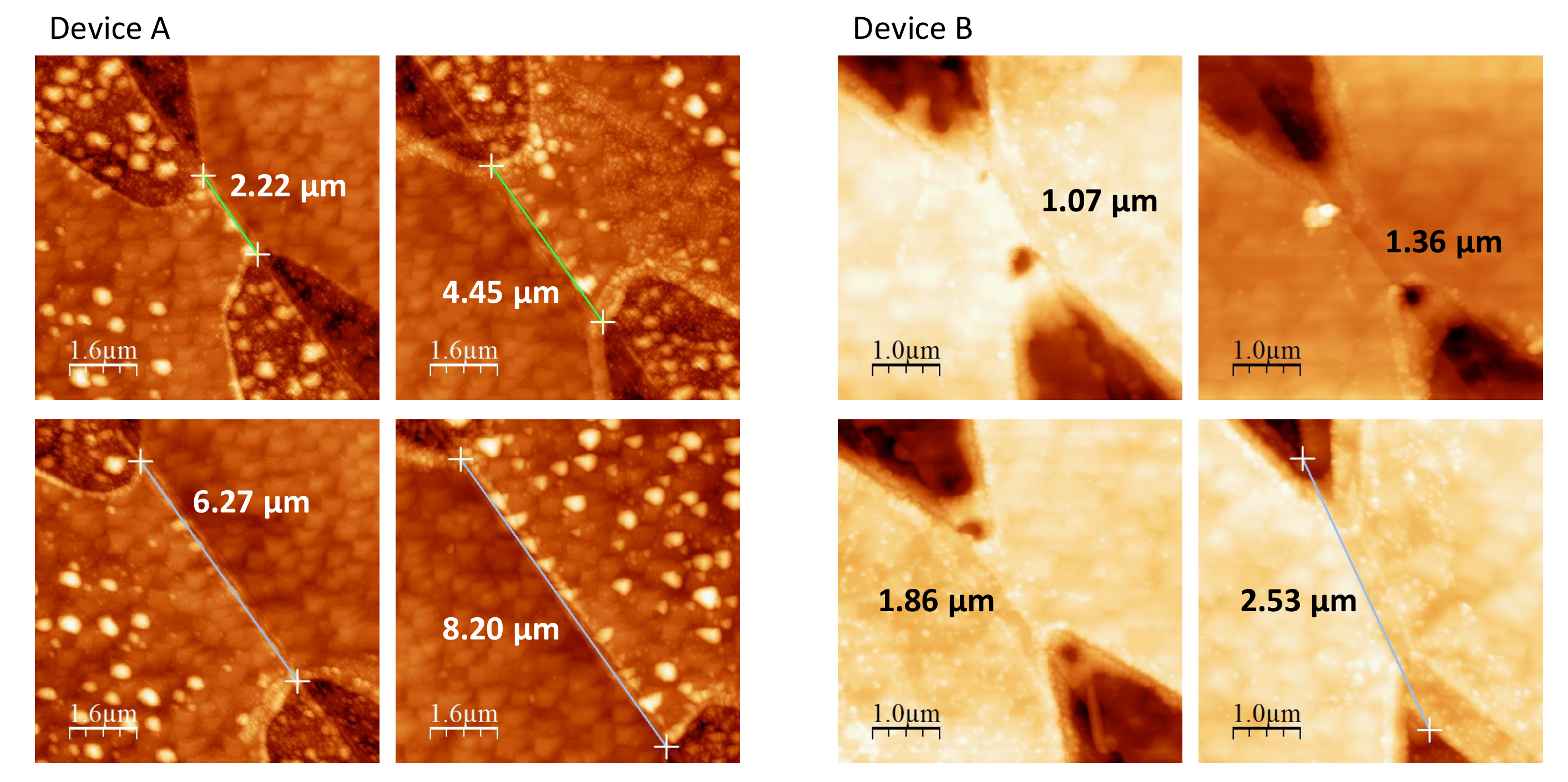}
\vspace{0in}
\caption{AFM micrographs of Devices A and B. A faint line across the constriction is the Ti gate boundary.}
\label{SAFM}
\end{figure}

Images of constrictions with the gate boundaries are shown in Fig.~\ref{SAFM}. Numbers indicate lithographical width of constrictions along gate boundaries.

\section{2DEG preparation}

We found that conditions of LED illumination have a great impact on the width and electrostatic control of the QHFm transition. Illumination of a sample with a red LED at $\sim$10~K results in a very wide ($0.5-0.8$~T) QHFm transition which has a position that is sensitive to the gate voltage ($0.3-0.4$~T per 100~V on a back gate voltage). Similar results were obtained by illuminating a sample with a green LED at low temperatures $\sim200$~mK. Illumination with a green LED at high temperatures ($\sim$10~K) results in a 2D gas with $\sim30\%$ higher carrier density and narrow ($0.1-0.3$~T) QHFm transition with a position almost insensitive to the applied gate voltage.

The optimal QHFm transition width and gate sensitivity was achieved by illuminating samples with a green LED at low temperatures and subsequent heating to 1~K, where after 2-4~hours the 2D gas relaxed into an intermediate state with a $0.2-0.4$~T - wide QHFm transition and $0.1-0.2$~T/100~V transition control. Thus, prepared 2D gases vary slightly between cooldowns for the same sample and between different samples.

\begin{figure}[h]
\centering\includegraphics[width=0.75\columnwidth]{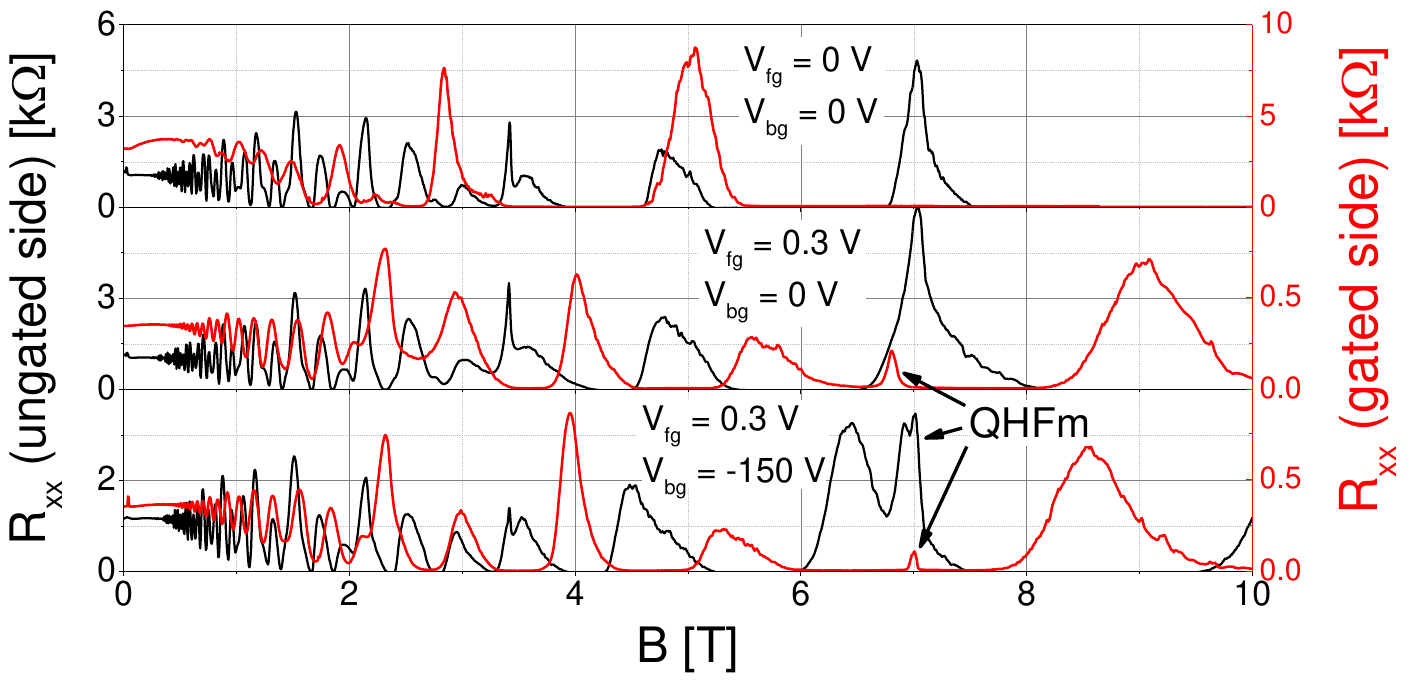}
\vspace{0in}
\caption{After cooldown densities under gate and outside the gate differ by a factor of 2-3 due to surface Fermi level pinning by the gate. In order to align densities one needs to apply a high voltage on the front gate. After aligning densities and placing $\nu=2$ in the vicinity of the QHFm transition one can observe QHFm transitions on both ungated and gated sides.}
\label{Sdensity}
\end{figure}

The Ti front gate, evaporated directly on the CdTe surface, is found to modify surface pinning potential and reduce electron density under the gate by a factor of 2, see Fig.\ref{Sdensity}. Modified surface pinning potential causes different dopant ionization profiles and, consequently, different profile of the electron wavefunction within the quantum well. This difference allowed us to adjust the transition field $B^*$ at zero gate voltage by varying conditions of the LED illumination during a cooldown. Sharp peaks near 3.5~T and 7~T are QHFm transitions due to Landau level crossings. Adjusting front and back gate voltages we can position the 7~T QHFm transition between $\ket{1\downarrow}$ and $\ket{0\uparrow}$ states within the $\nu=2$, as shown in the middle and bottom panels in Fig.~\ref{Sdensity}.

\section{Characterization of local heating by excitation current}

\begin{figure}[h]
\centering\includegraphics[width=0.5\columnwidth]{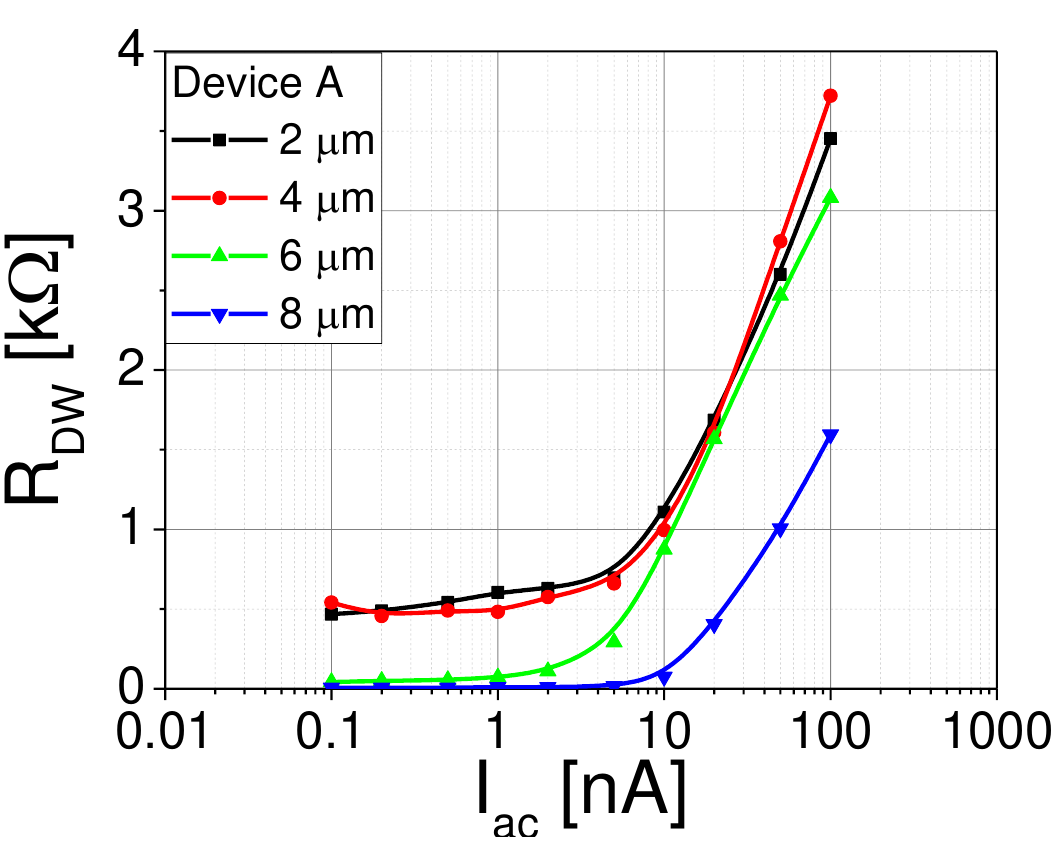}
\vspace{0in}
\caption{Dependence of $R_{DW}$ for different channel lengths on excitation current at a base temperature $T=27$ mK.}
\label{Sidep}
\end{figure}

Current dependence of $R_{DW}$ measured at base $T<30$~mK is shown in Fig.\ref{Sidep}. Saturation of $R_{DW}$ at low currents indicates that for excitation currents $I_{ac}<1$~nA Joule heating is negligible.

\section{Modeling longitudinal resistance in the presence of a domain wall}

\begin{figure}[h]
\centering\includegraphics[width=\columnwidth]{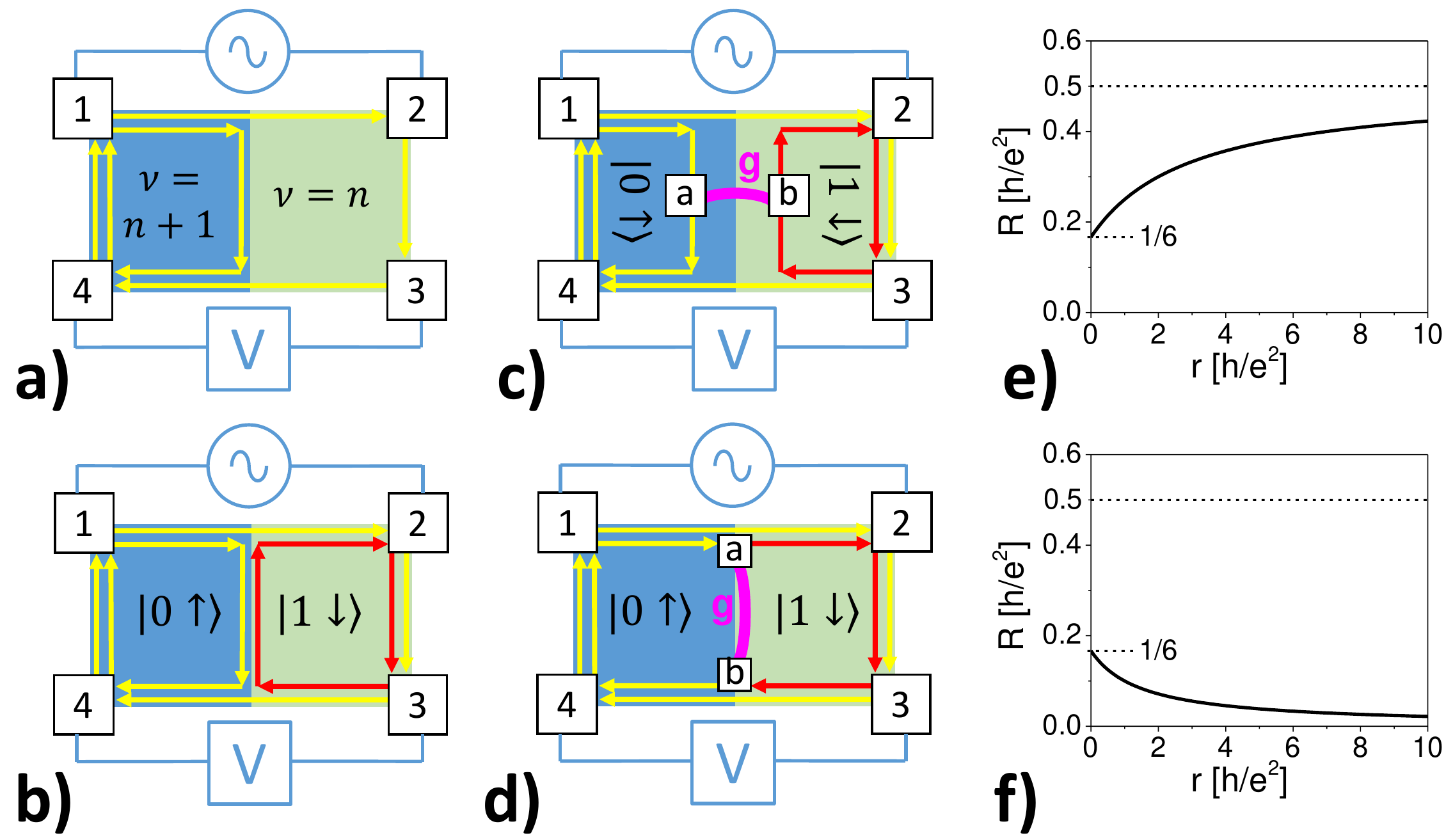}
\vspace{0in}
\caption{(a) Edge states in a sample with different filling factors. (b) Edge states at the domain boundary in the absence of inter-edge scattering. (c) The same with inter-edge scattering $g$ or (d) formation of a hDW with conduction $g$. (e,f) Dependence of longitudinal resistance $R$ on conduction $g=1/r$ calculated for models (c,d).}
\label{Smodel}
\end{figure}

Multi-terminal transport in the quantum Hall effect regime can be accurately modeled within a Landauer-B\"uttiker formalism. First, let us calculate the longitudinal resistance in a sample with variable electron density, across a chiral edge state formed at the density boundary, see Fig.\ref{Smodel}(a). We have a sample with four contacts and two areas of different filling factors ($\nu=n$ and $\nu'=n+1$). In this case if we would pass current $i$ from source {contact \bf 1} to drain {contact \bf 2} the measured voltage drop between contacts {\bf 3} and {\bf 4} can be found from solving a set of Kirchhoff's equations:

\begin{equation}
\begin{cases}
(n + 1) * g_0 * V_4 - (n + 1) * g_0 * V_1 + i = 0\\
V_2 = 0 \\
n * g_0 * V_2 - n * g_0 * V_3 = 0 \\
n * g_0 * V_3 + g_0 * V_1 - (n + 1) * g_0 * V_4 = 0
\end{cases}.
\end{equation}

Here $g_0$ is the quantum conductance $\frac{e^2}{h}$. The resistance across the chiral edge state $R_{ch}$ is 

\begin{equation}
R_{ch}=\frac{V_4-V_3}{i} = \frac{1}{n(n+1)}\frac{1}{g_0} = \frac{1}{n(n+1)}\frac{h}{e^2}.
\end{equation}

This result was obtained for one magnetic field direction. Magnetic field reversal would flip the direction of propagation of the edge states. By rewriting the system of equations one can find potentials of contacts {\bf 3} and {\bf 4}. They are the same $V_3 = V_4 = \frac{i}{g_0 n}$. Thus, in reversed magnetic field $R_{ch} = 0$.

Now we turn to the modeling of resistance in the presence of a hDW at $\nu=2$. First, we consider the case of two non-interacting counterpropagating edges with no inter-scattering and no equilibration between $\nu=2$ and $\nu=1$ edge states, Fig.\ref{Smodel}(b). The solution is $R_{DW}=\frac{V_4-V_3}{i} = \frac12\frac{h}{e^2}$=12.9~k$\Omega$, independent of field direction.

Analyzing resistivity of the domain wall constriction, we consider two possibilities. The first is the inter-edge scattering between counterpropagating edge states along the gate boundary. In this case, we parametrize the domain wall resistance by a finite conductivity $g=1/r$ between edge channels, Fig.\ref{Smodel}(c). The system of Kirchhoff's equations for this case is:
\begin{equation}
\begin{cases}
2 * g_0 * V_4 - 2 * g_0 * V_1 + i = 0\\
g_0 * V_1 - g_0 * V_a + g * (V_b - V_a) = 0 \\
g_0 * V_3 - g_0 * V_b - g * (V_b - V_a) = 0 \\
V_2 = 0 \\
2 * g_0 * V_2 - 2 * g_0 * V_3 = 0 \\
g_0 * V_a + g_0 * V_3 - 2 * g_0 * V_4 = 0
\end{cases},
\end{equation}
and $R_{DW}=\frac{r+1}{2r+6}\frac{h}{e^2}$, Fig.\ref{Smodel}(e). While the value $1/6<R_{DW}<1/2$ depends on $r$, it is inconsistent both with the value of resistance and with dependence of resistance on the length of the hDW channel observed experimentally. Indeed, interedge scattering would mostly depend on the width of the channel, and will not exhibit exponential dependence on its length. 

A hDW formed from fully hybridized counter-propagating edges is modeled as a single channel with conductivity $g=1/r$ connecting $\nu=2$ edge states on the opposite sides of the sample, Fig.\ref{Smodel}(d). In this case Kirchhoff's equations are:

\begin{equation}
\begin{cases}
2 * g_0 * V_4 - 2 * g_0 * V_1 + i = 0\\
g_0 * V_1 - g_0 * V_a + g * (V_b - V_a) = 0 \\
V_2 = 0 \\
2 * g_0 * V_2 - 2 * g_0 * V_3 = 0 \\
g_0 * V_3 - g_0 * V_b - g * (V_b - V_a) = 0 \\
g_0 * V_b + g_0 * V_3 - 2 * g_0 * V_4 = 0
\end{cases},
\end{equation}
and in this case $R_{DW}=\frac{1}{4r+6}\frac{h}{e^2}$, Fig.\ref{Smodel}(f). Experimentally observed resistance is consistent with this picture and the theoretical model of conduction through in-gap states is discussed in Section S9. By substituting $r\rightarrow\infty$, this case corresponds to an insulating hDW and is reduced to the usual quantum Hall case with $R_{xx}=0$. This situation was observed for long hDW, longer than 6-8~$\mu$m.

We note that in the case of a highly conducting hDW, $r\rightarrow0$, we get $R=\frac16\frac{h}{e^2}$ as in the previous case, Fig.\ref{Smodel}(c,e). For models Fig.\ref{Smodel}(c,d) with $r\rightarrow0$ it is easy to show that the corresponding systems of equations are the same. Physically it means that points {\bf a} and {\bf b} have the same potential and can be merged together on Figs.\ref{Smodel}(c,d). It's easy to show that for all hDW models $R_{DW}(B)=R_{DW}(-B)$, reflecting the fact that hDWs are symmetric under B-field inversion.

\section{Dependence of conductance fluctuations on $\Delta B^*$}

\begin{figure}[h]
\centering\includegraphics[width=\columnwidth]{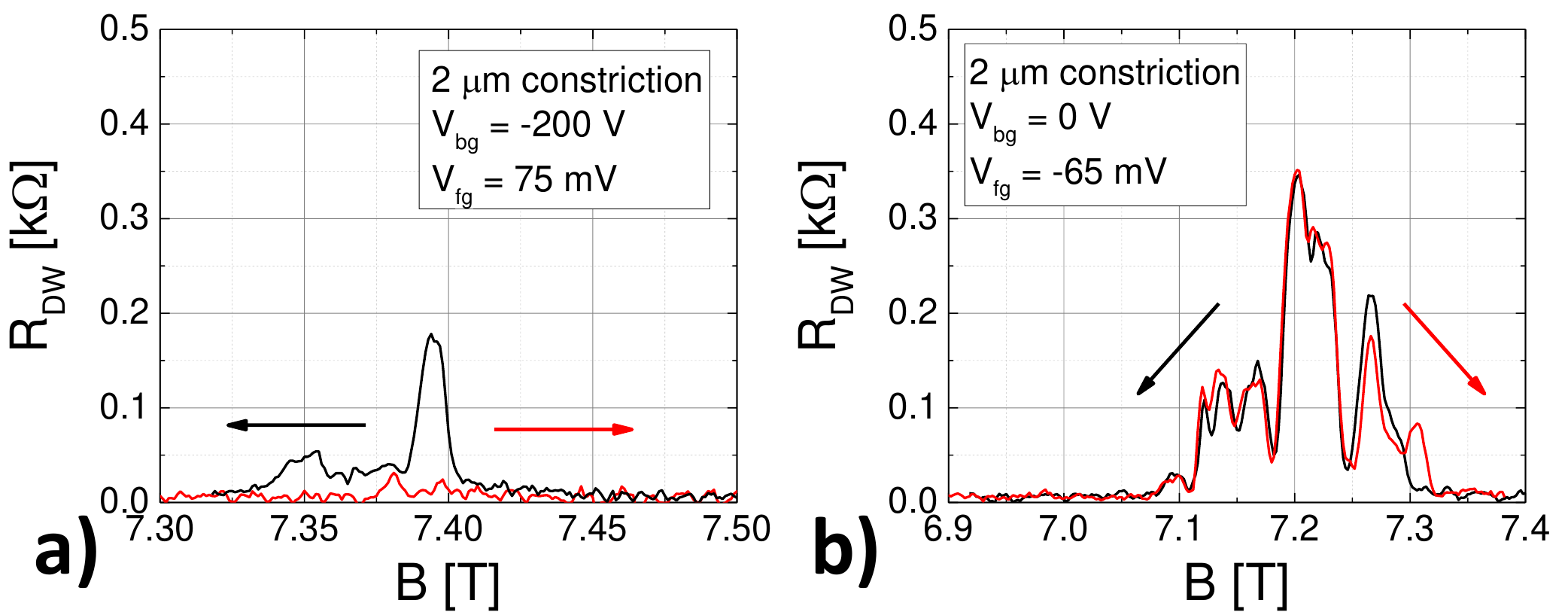}
\vspace{0in}
\caption{Mesoscopic resistance fluctuations for 2~$\mu$m channels are shown for (a) $\Delta B^*=0$ and (b) $\Delta B^*=0.2$~T. For each temperature point consecutive $B$-scans in both field directions were recorded.}
\label{Shyst}
\end{figure}

Separation of QHFm transitions in gated and ungated regions $\Delta B^*$ reflects the value of the s-d exchange gradient near the gate boundary and, as a result, positions of ferromagnetic domains. For $\Delta B^*=0$ there is no s-d exchange gradient and domains are randomly formed within the 2D plane. Narrow field sweeps within $7.3~T<B<7.5~T$ range across the QHFm transition results in the formation of different domain configurations, different conduction paths, and different patterns of conduction fluctuations, Fig.~\ref{Shyst}(a). Often there is no domain wall formed in the vicinity of the constriction, in this case no conduction is observed as shown for the up-sweep in (a). In contrast, for $\Delta B^*=0.2$~T the gate-induced s-d exchange gradient stabilizes the domain wall position and conducting channels are always formed. The conduction channel is well defined and the resistance fluctuation pattern is reproducible over multiple field sweeps, Fig.~\ref{Shyst}(b).

\section{Time evolution of mesoscopic fluctuations}

\begin{figure}[h]
\centering\includegraphics[width=\columnwidth]{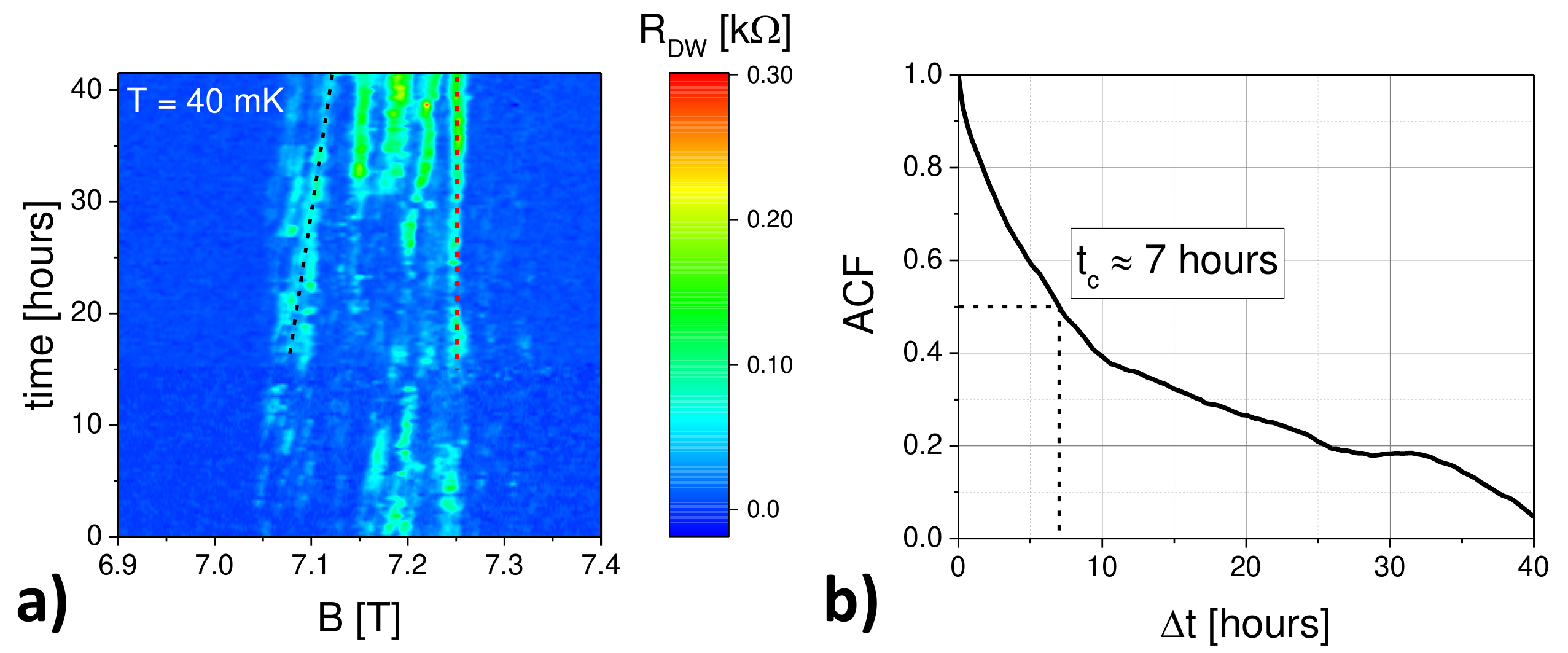}
\vspace{0in}
\caption{Time evolution of resistance fluctuations in a 2~$\mu$m channel is plotted in the color plot for $\Delta B^*=0.2$~T and $T=40$~mK. Data was recorded for $B=6.9~T\rightarrow7.4~T$ sweeps. (b) Resistance auto-correlation function as a function of time offset $\Delta$t.}
\label{Stime_dep}
\end{figure}

Even for large QHFm transition separation $\Delta B^*=0.2$~T and at the lowest $T<30$~mK there is a slow change in the pattern of resistance fluctuations with time, Fig.~\ref{Stime_dep}. A characteristic time scale for the pattern change is $\sim7$~hours, as determined from a half width at the half height of the autocorrelation function $F(\Delta t) = \left<R_{DW}(t)R_{DW}(t+\Delta t)\right>$. Most likely the conduction path and the fluctuation pattern are affected by gate voltage-induced slow motion of localized charges in the vicinity of the conduction channel.

\section{Dependence of helical channel conductance on the position of $B^*$ within the $\nu=2$ plateau}

\begin{figure}[h]
\centering\includegraphics[width=\columnwidth]{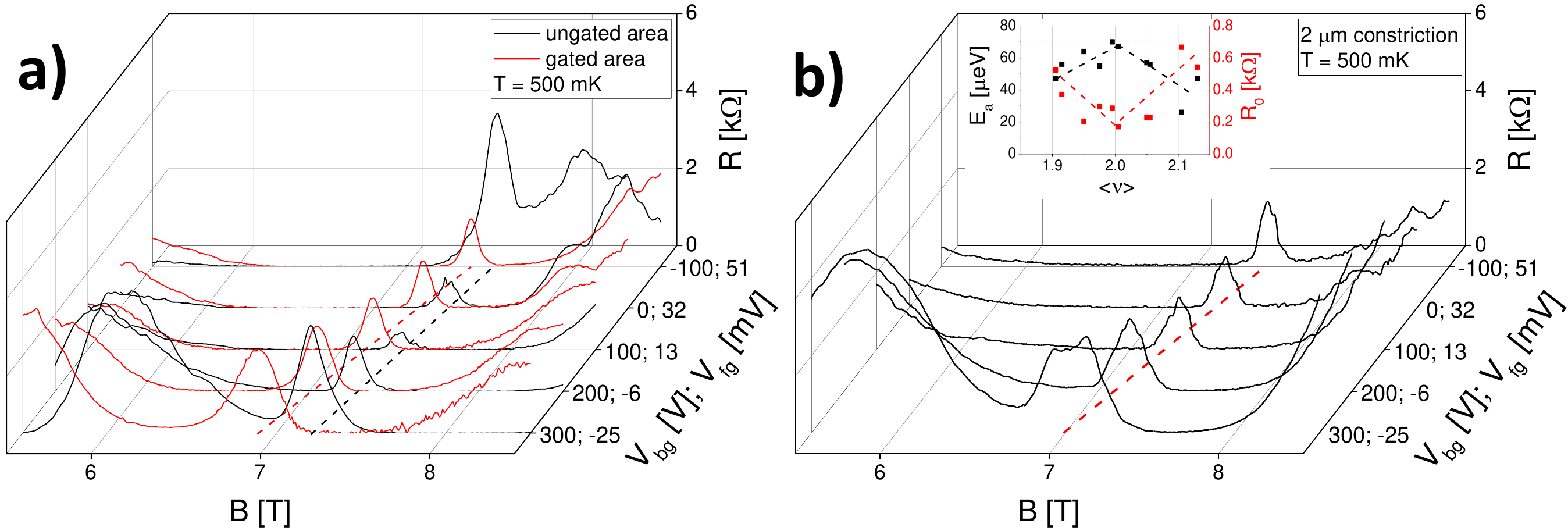}
\vspace{0in}
\caption{$R_{DW}$ in the vicinity of the $\nu=2$ QHE state is plotted for gated and ungated regions (left) and across a 2~$\mu$m constriction for different front gate $V_{fg}$ and back gate $V_{bg}$ voltages at $T=500$~mK. Here position of the QHFm transition $\langle B^*\rangle$ is shifting relative to the center of the $\nu=2$ plateau in both gated and ungated regions. In the inset the value of saturation resistance $R_0$ and activation energy $E_a$ are plotted as a function of a filling factor of $\langle B^*\rangle$, where $R_0$ and $E_a$ are extracted from the constant + activation function fits to the temperature dependence of QHFm transition peaks.}
\label{nudep}
\end{figure}

The value of the maximum conductivity of the channel (which corresponds to the maximum of $R_{DW}$) formed between states with opposite spin polarization depends not only on the length of the channel and QHFm separation $\Delta B^*$, but also on the position of the QHFm transition within the $\nu=2$ plateau. In Fig.~\ref{nudep} we simultaneously change density in gated and ungated regions and sweep the QHFm in the channel $\langle B^*\rangle=(B^*_{gated}+B^*_{ungated})/2$ across the $\nu=2$ plateau while keeping $\Delta B^*$ approximately constant. Magnetoresistance in the gated and ungated regions is plotted in the left plot, and across the 2~$\mu$m constriction in the right plot, Fig.\ref{nudep}. In the inset, the resistance saturation value $R_0$ and activation energy $E_a$ are extracted from the temperature dependence of $R_{DW}$. It is clear that extrema of $R_0$ and $E_a$ depend on the position of $\langle B^*\rangle$ within the $\nu=2$ plateau, with the minimum $R_0$ and the maximum $E_a$ occur at $\nu=2$. The data discussed in the main text is taken for $\langle B^*\rangle$ placed close to the center of the $\nu=2$ plateau in both gated and ungated regions.

\section{Temperature dependence of resistance fluctuations and the phase coherence length}

\begin{figure}
\centering\includegraphics[width=0.9\columnwidth]{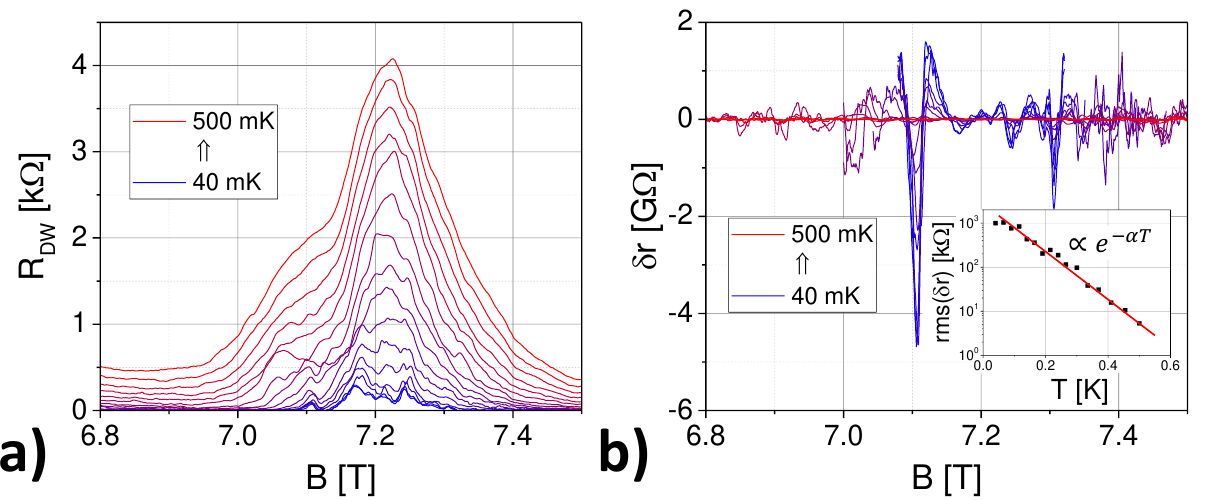}
\vspace{0in}
\caption{(a) Resistance $R_{DW}$ across 2~$\mu$m constriction at different temperatures. (b) Corresponding fluctuations $\delta r$ of resistance r along the gate boundary. Inset shows standard deviation of resistance across a 2~$\mu$m constriction.}
\label{STdep}
\end{figure}

Temperature dependence of resistance fluctuations across a $2\ \mu$m constriction is shown in Fig.~\ref{STdep}a, where the magnetic field was swept in a narrow range near the the QHFm transition (6.8-7.5~T) in order to preserve the fluctuation pattern. The channel resistance $r$ can be calculated from the measured resistance $R_{DW}$ using Landauer-B\"uttiker formalism discussed earlier:
\begin{equation}
r=\frac{1-6R_{DW}}{4R_{DW}},
\end{equation}
where both $r$ and $R_{DW}$ are expressed in units of $h/e^2$. Fluctuations of the measured resistance $\delta R_{DW}$ are obtained by subtracting a smooth background from the resistance $R_{DW}$, and fluctuations of the resistance of the conducting channel $\delta r$ are calculated as
\begin{equation}
\delta r=\frac{dr}{dR_{DW}}\delta R_{DW}=-\frac{1}{4}\frac{\delta R_{DW}}{R^2}.
\end{equation}
Extracted fluctuations of channel resistance are plotted in Fig.~\ref{STdep}b for a wide temperature range. In the inset rms amplitude of $\delta r$ is plotted as a function of temperature. From exponential decay of rms($\delta r$) with temperature $rms\propto e^{-T_0/T}=e^{-L/l_{\phi}(T)},\ T_0=80$~mK, we estimate that phase coherence length $l_{\phi}$ exceeds $\approx800$~nm below 100~mK for $L\approx1~\mu$m. Thus the phase coherence is preserved over the length of the channel.

\section{Modeling of domain wall conduction in the {QHFm} regime}

\begin{figure}[b]
\includegraphics[width=0.4\columnwidth]{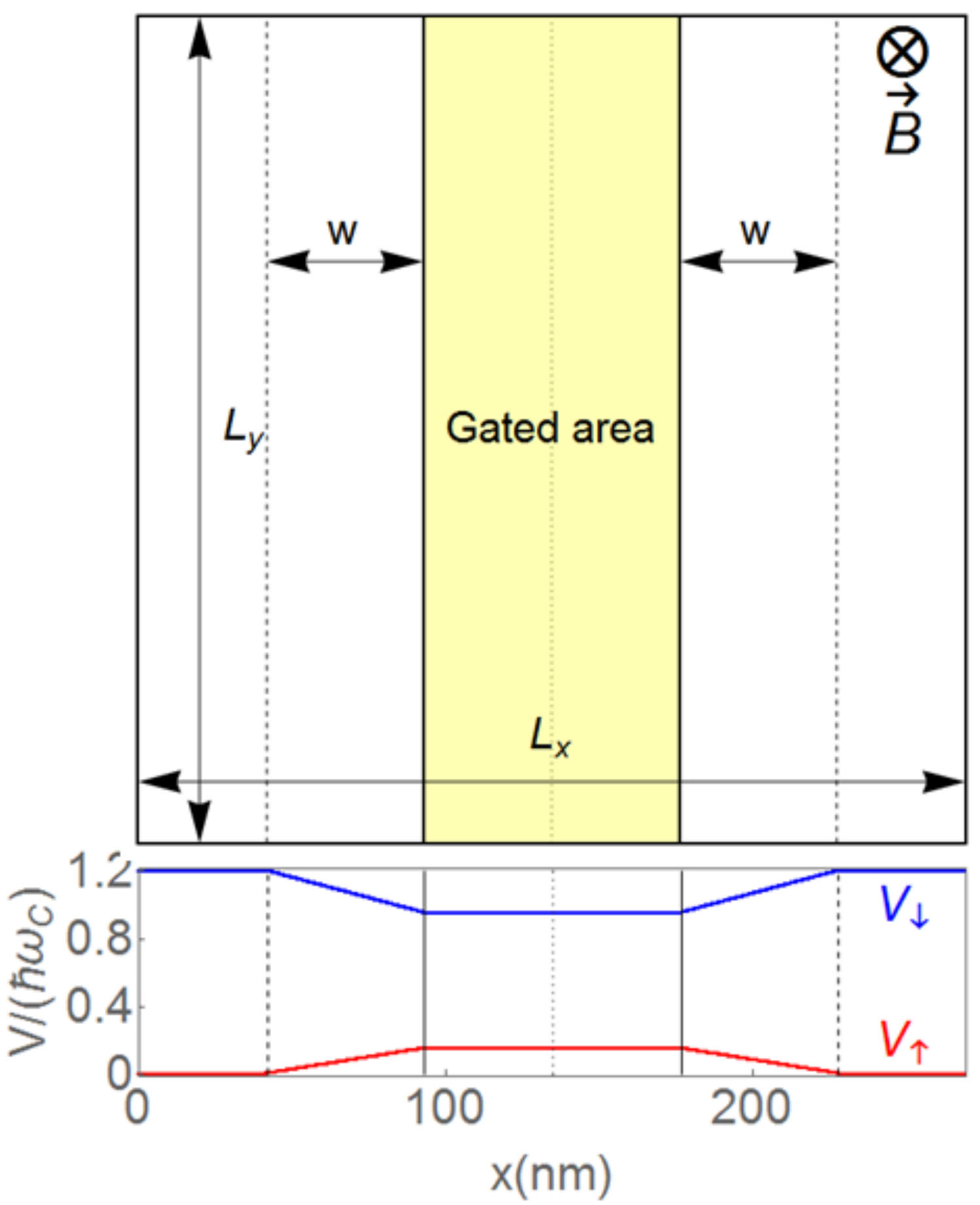}
\caption{Schematic view of the simulated system and the spin dependent potential due to gate voltage.}
\end{figure}

In order to model our system, we consider $N$ electrons confined to a $L_x\times L_y$ rectangle, subjected to a magnetic field $B=-B \hat e_z$. We take $N=[\nu L_x L_y /2\pi\ell^2]$, where $\ell$ is the magnetic length and $\nu=2$ is the filling factor.

\begin{eqnarray}
H_s&=&\sum_{i}\big[ \frac{1}{2m^*} \left({\bf p}_i+\frac{e \bf{A}}{c}
\right)^2 +\frac{\beta_R}{\hbar} \left({\bf p}_i
\frac{e \bf{A}}{c}
\right)\times\boldsymbol{\sigma}_i \big.\nonumber\\
&+&\big. V_G ({\bf r})
\sigma_{z,i}+V_{{\rm{imp}}}({\bf r}) +V_{{\rm{fl}}} ^{{\rm{mag}}}
({\bf r}) \sigma_z\big]\nonumber\\
&+&\frac{e^2}{2\epsilon_r
}\sum_{i,j}{\frac{1}{|{\bf{r}}_i-\bf{r}_j|}}
\end{eqnarray}

Here $m^*=0.1 m$ is the effective electron mass in CdTe and $\beta$ is the Rashba constant. The spin dependent potential $V_G$ mimics variation of the Zeeman energy across the sample as a result of applied gate voltage. We consider a remote impurity potential $V_{{\rm{imp}}}({\bf r})=\sum_{i}^{N_i} w_i\exp[-({\bf{r}}-{\bf{r_i}})^2/d^2]$, where the number of impurities $N_i=N$ and $r_i$'s denote the position of the randomly placed impurities in the doping layer and $w_i \in [-W, W]$. Surface roughness (see Fig.~6(e) of the main text) translates into a curvy profile of the quantum well, and as a consequence, into the deviation of magnetic field orientation. This deviation causes fluctuations of Mn spin orientation from the z-direction. In order to model this effect of surface roughness we introduce the spin dependent random potential $V_{\rm{fl}}^{\rm{mag}}({\rm{r}})=\sum_i u_i\exp[-({\bf{r}}-{\bf{r}}_i)^2/{b^2} ]\sigma_z$. We choose $W=8$~meV, $d=40$~nm, and $u=15$~$\mu$eV and $b=150$ nm. Parameters for remote dopants are chosen to be consistent with the electron mobility that has been measured experimentally ($\mu=30,000$~cm/Vs at $B=0$). The electron-electron interaction is taken into account using the Hartree-Fock approximation. The self-consistent procedure is done in the basis set of five orbital Landau states, each with two spin projections. In our numerical procedure, the spin-dependent potential and random impurities are chosen to be symmetric with respect to the reflection about a line parallel to the $y$-axis that bisects $L_x$; $V_G(x,y)=V_G(L_x-x,y)$ and $x_{N/2+i}=L/2-x_i$, $y_{N/2+i}=y_i$. Periodic boundary conditions are used in both $x$ and $y$ directions. The Hartee-Fock procedure reduces the Hamiltonian to a non-diagonal and non-local effective single particle form \cite{TheBook}.

This model yields two counterpropagating edge channels experiencing avoided crossing due to the spin-orbit gap. Impurities provide states in the gap mediating the conduction in short channels. We compute the conductance of our finite system using a Green's function approach \cite{TranspBook}. Knowing the single particle Hartree-Fock and impurity potential, we discretize the problem on a lattice of $N_x\times N_y$ points. We place our leads in the channels separated by $L_y/2$. The Hamiltonian describing the system with leads is given by
\begin{equation}
H_t=H_s+H_1+H_2+V_{1s}+V_{2s}~,
\end{equation}
where $H_i$ describes the lead, $V_{is}$ is the coupling between lead and the localized electron states in the domain wall area ($i=1,2$ label the lead). The conductance is given by
\begin{equation}
G=\frac{e^2}{h}{\rm{Tr}} \left(\hat\Gamma_1 \hat{\mathcal G_{R}}\hat\Gamma_2 \hat {\mathcal G_{A}} \right)~,
\label{eqG}
\end{equation}
where $\mathcal G_{R/ A}$ denotes the retarded (advanced) Green's function of the interacting electron gas, $\hat {\mathcal G}_{A,R}=[(E\pm i\eta)\hat I-\hat H]^{-1}$, $E$ is the energy (we take $E=E_F$), $\hat \Gamma_i=i(\hat\Sigma_i^{R}-\hat\Sigma_i^{A})$ are the coupling matrices, and the contact retarded and advanced self-energies $\hat\Sigma_i^{R}$ and $\hat\Sigma_i^{A}$ are given by
\begin{eqnarray}
\hat\Sigma_i^{R}&=&V_{is}^{\dagger}\left[ \left(E+i\eta\right)\hat I-\hat H_{i}\right]^{-1}\hat V_{is}\\
\hat\Sigma_i^{A}&=&V_{is}^{\dagger}\left[ \left(E-i\eta\right)\hat I-\hat H_{i}\right]^{-1}\hat V_{is}~.
\end{eqnarray}

We compute the conductance using (\ref{eqG}) and extract the conductivity of the hDW $\sigma_{yy}$. When both magnetic and remote impurities are present, the averaged conductivity for five realizations of disorder is found to be $1/r=\sigma_{yy}=0.146\pm0.023~{\frac{e^2}{h}}$. If magnetic fluctuations are ignored, we obtain $\sigma_{yy}=0.105\pm0.018~{\frac{e^2}{h}}$. The calculated value $\sigma_{yy}=0.146~{\frac{e^2}{h}}$ corresponds to the channel resistance $r=1/\sigma_{yy}=177$~k$\Omega$ or $R_{DW}=0.77$~k$\Omega$. This value of $R_{DW}$ is in good agreement with the measured resistance $R_{DW}=0.66-0.87$~$k\Omega$, suggesting that the model captures the essential physics of conduction in the channels formed along domain walls. In-gap states naturally provide conduction channels for electrons propagating in both directions. Therefore, the system yields resistivity $R_{DW}$ symmetric under magnetic field direction reversal, in agreement with experiment.

\end{document}